\documentclass[10pt,letterpaper,twocolumn]{article} 

\usepackage{ol2}
\usepackage{hyperref}
\usepackage{amsmath}
\usepackage{setspace}
\usepackage[font=small]{caption}
\begin{document}

\twocolumn[ 

\title{Optimization of gain in traveling-wave optical parametric amplifiers by tuning the offset between pump- and signal-waist locations}
\vspace{-5mm}
\author{Gideon Alon,$^{1}$ Oo-Kaw Lim,$^{1}$ Amar Bhagwat,$^{1}$ Chao-Hsiang Chen,$^{1}$ Muthiah Annamalai,$^{2}$ \\ Michael Vasilyev,$^{2}$ and Prem Kumar$^{1,*}$}
\address{
$^{1}$Center for Photonic Communication and Computing, EECS Department,  Northwestern University, \\ 2145 Sheridan Road, Evanston, IL 60208, USA \\
$^{2}$Department of Electrical Engineering, University of Texas at Arlington, Arlington, TX  76019, USA \\
$^{*}$Corresponding author: kumarp@northwestern.edu
}
\vspace{-3mm}
\begin{abstract}We present experimental demonstration and modeling of the optimization of a phase-sensitive optical parametric amplifier by tuning the relative position between the pump- and signal-beam waists along the propagation direction. At the optimum position, the pump beam focuses after the signal beam, and this departure from co-located waists increases with increasing pump power. Such optimization leads to more than 3\,dB improvement in the measured de-amplification response of the amplifier.\end{abstract}

\vspace{-7mm}
]
\noindent
Optical parametric amplifiers\,(OPAs) are ubiquitous these days both in pure research and in industrial applications. Hence, it is important to adjust these OPA systems for optimum matching with the input signal and detector modes. In squeezing experiments, a matched local-oscillator\,(LO) that properly extracts the distorted spatiotemporal squeezed mode can optimize the detected amount of squeezing\cite{KIM1994}. In the same spirit, we report here on the performance optimization of an OPA-based system by adjusting the longitudinal offset between the waist locations of the pump and signal beams. We find that the optimal offset is significantly different from zero, contrary to conventional wisdom.

\begin{figure}[htb]
\centerline{\includegraphics[width=5.5cm]{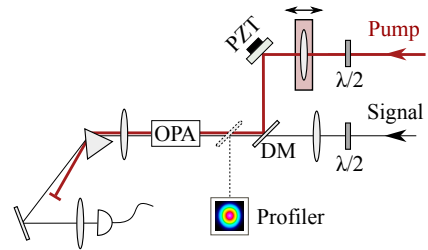}}
\caption{(Color online) Experimental configuration of the OPA. PZT: piezoelectric transducer, DM: dichroic mirror, $\lambda/2$: half-wave plates.}
\label{fig1}
\end{figure}

In Fig.\,\,\ref{fig1} we show the key elements of the experimental setup. Light from a telecom-band (1560\,nm) continuous-wave (CW) distributed-feedback (DFB) laser is fed into a pulse carver that outputs an 8\,MHz train of 160\,ps flat-top pulses. These pulses are subsequently amplified by a series of erbium-doped fiber amplifiers (EDFAs) to reach peak powers of $>$2\,kW. This pulsed vertically polarized light is then collimated into a free-space beam and focused into a 1-cm-long periodically-poled potassium-titanyl-phosphate (PPKTP) crystal for second-harmonic generation (SHG). Conversion efficiencies of $>$65\% produce a pulse train at 780\,nm with $>$1\,kW peak power that is then used to pump a 3-cm-long PPKTP-based OPA stage. A small signal to be amplified by the OPA is tapped from the main beam prior to the SHG stage, because after the SHG stage the spatiotemporal mode at 1560\,nm is severely distorted due to the high conversion efficiency. Both crystals (Raicol Crystals Ltd.) are poled for type-0 interaction (i.e., all fields are co-polarized). The OPA stage is operated as a degenerate OPA (DOPA). Each of the crystals is held at the phase-matching temperature with better than 10\,mK stability. The pump and signal beams are combined on a dichroic mirror (DM) and are directed into the OPA crystal. The focusing lens of the pump beam is mounted on a micrometer-controlled translation stage, allowing for fine adjustment of the pump-waist location inside the crystal along the propagation direction. A tap is placed before the crystal, which directs both beams to a spatial profiler (DataRay Inc., model BMS2-CM4-IGA) capable of measuring beam waists down to ${\rm 1\,\mu m}$ with ${\rm 0.1\,\mu m}$ accuracy and the focal-point positions with ${\rm \pm1\,\mu m}$ repeatability. The profiler is placed near the focal plane and is used to determine the locations and values of both Gaussian beams' waists in free space prior to measuring the OPA gains. The pump and signal waists are measured to be $(a^{px}_0, a^{py}_0, a^{sx}_0, a^{sy}_0)=(27.1, 25.8, 36.8, 38.6)\,{\rm\mu m\,\pm 2 \,\mu m}$ with $a_{0}$ being the $1/e$ intensity radius, which display an approximate $\sqrt{2}$ relationship between the signal and pump waists. Parametric gain and de-gain values are then measured for various positions of the pump lens via direct detection of the signal beam with a pulse-resolving detector having a bandwidth of 12.5\,GHz. The distance along the propagation direction between the waists of the pump and signal beams in the crystal, referred to as $z$-offset, is inferred by multiplying the translation-stage micrometer readings by the refractive index of the crystal. A piezo-driven mirror controls the relative optical phase between the signal and pump beams. Throughout data collection, the signal beam's waist is not adjusted. 

To model the expected gain/de-gain performance, an implementation of the beam-propagation method using fast-Fourier-transform (BPM-FFT) is employed. In this method, the continuous-wave field inside the crystal is modeled in discrete symmetrized steps of length $h$. An initial free-space propagation step of length $h/2$ is followed by diffraction-free nonlinear interaction over a step of length $h$, which is followed further by a free-space step of length $h/2$. The free-space propagation part of the degenerate phase-matched nonlinear paraxial Helmholtz equation for the signal field ${\cal E}$,
\begin{equation}
\frac{d{\cal E}(\bar{\rho},z)}{dz} + \frac{1}{2ik}\nabla^2_{\perp}{\cal E}(\bar{\rho},z) = K{\cal E}^{*}(\bar{\rho},z){\cal E}_{P}(\bar{\rho},z)
\label{eq:nonlin.helmholtz}
\end{equation}
with $K=i\omega_{s}\chi^{(2)}/n_{s}c$ and Gaussian pump field ${\cal E}_P$, is solved in the Fourier domain by a simple multiplication, where the operator $\nabla^{2}_{\perp}$ reduces to a factor $-(k_x^2+k_y^2)$ with $k_j$ being the angular spatial frequency, $j=x,y$. The nonlinear part is solved at each iteration step by using the thin-crystal solution (i.e., point-by-point amplification)
\begin{align}
{\cal E}(\bar{\rho},z_{n+1}) = {\cal E}(\bar{\rho},z_{n}+h_1)\cosh[\kappa(\bar{\rho},z_n&+h_1)h]\\
+\,ie^{[i\varphi_{P}(\bar{\rho},z_n+h_1)]}\sinh[\kappa(\bar{\rho},z_n+h_1)h]&{\cal E}^{*}(\bar{\rho},z_{n}+h_1),\nonumber
\label{eq:bpm.nonlin.sol}
\end{align}
where $\chi^{(2)}=2\,d_{\rm eff}$, $h_1=h/2$, $\varphi_{P}(\bar{\rho},z)=\arg[{\cal E}_{P}(\bar{\rho},z)]$ and
$\kappa(\bar{\rho},z)=\frac{2\omega_{s} d_{\rm eff}}{n_{s}c}|{\cal E}_{P}(\bar{\rho},z)|$.
The residual pump-signal phase fluctuations $\Delta \theta$ that average out the phase-sensitive gain and de-gain performance of the OPA are included by writing the gain as 
\begin{equation}
G\left(\Delta \theta\right)=G\cos^2 \Delta \theta + G^{-1}\sin^2 \Delta \theta\,.
\label{eq:nonlin.gain}
\end{equation}

\begin{figure}[htb]
\centerline{\includegraphics[width=7.6cm]{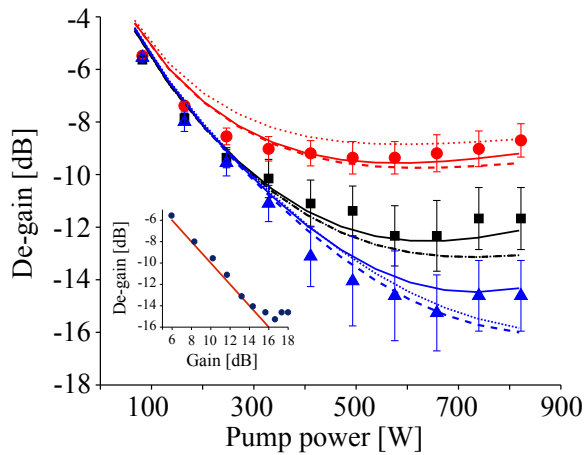}}
\caption{(Color online) Gain as a function of pump power for different $z$-offsets. Red circles show experimental results for $z$-offset of +5.5\,mm (signal waist is downstream from the pump waist). Black squares are data for $z$-offset $=0$\,mm and blue triangles are data for $z$-offset $=-3.7$\,mm. Results from the model: dotted lines are obtained by scanning the signal waist, dashed lines by scanning the pump waist, and solid lines by scanning the pump waist while including phase fluctuations with $\Delta \theta = 0.65^{\circ}$. The $d_{\rm eff}$ parameter is used to scale the horizontal axis of the modeled gain for best fit and is found to be 6.3\,pm/V. Inset shows de-gain as a function of gain for $z$-offset $=-3.7$\,mm. Straight line has 1:1 slope.}
\label{fig2}
\end{figure}
In Fig.\,\,\ref{fig2} we show the de-gain behavior as a function of the pump power. A clear advantage (i.e., greater degree of de-amplification) is found for choosing a $z$-offset of $-3.7$\,mm compared to co-located waists (i.e., zero offset). As expected, for larger offsets between the pump and signal waist locations, significant deterioration of the de-gain is also observed. Considering the confocal length $k_{0}a_{0p}^{2}$ of about 1\,cm in these measurements, the amount of offset required for optimal de-gain is substantial. We compare the experimental performance with predictions of the above BPM-FFT model for the case where the pump-waist location is scanned and the signal-waist location is kept fixed (similar to the case in our measurements) and for the case where the signal is scanned and the pump is kept fixed. 
\begin{figure}[htb]
\centerline{\includegraphics[width=7.6cm]{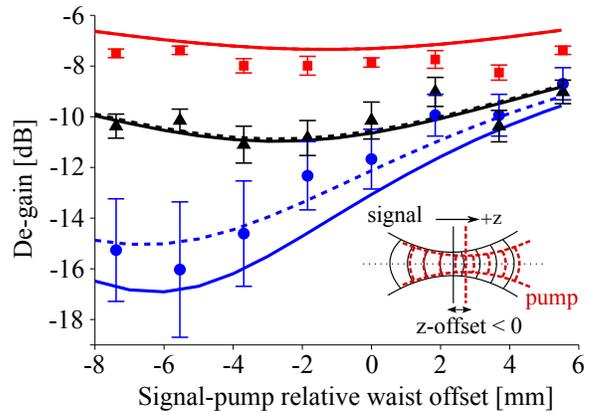}}
\caption{(Color online) De-gain dependence on $z$-offset for pump power of 165\,W (red squares), 330\,W (black triangles) and 820\,W (blue circles). Solid lines are results of simulations obtained by scanning the pump waist. Broken lines are similar to the solid lines but with inclusion of $\Delta\theta = 0.65^{\circ}$. Inset shows our sign convention of the $z$-offset along the propagation direction. Red broken line is the pump and black thin line is the signal. Negative offset is defined when the pump waist is located after the signal waist.}
\label{fig3}
\end{figure}
Figure\,\,\ref{fig3} shows the de-gain behavior as a function of the $z$-offset for various pump powers. The beam propagation is from left to right where negative offset means the signal's waist is located before the pump's in the propagation direction. Clear advantage for negative offsets is observed in both the experiments and models, where the departure from co-located waists increases at higher powers. At the highest pump power shown, with $z$-offset tuning, a de-gain enhancement of more than 3\,dB is observed over the case with zero $z$-offset. 

Gain-induced diffraction\cite{choi1997} in traveling-wave OPA distorts the signal wavefront so that some portions of it are in phase for gain while other portions get phased for de-gain. This behavior degrades the overall phase-sensitive amplifier (PSA) performance in direct-detection, particularly for the de-gain quadrature, because the measured signal power is integrated over the entire wavefront. Fine tuning of the $z$-offset finds the best spot where, in the more critical case of de-gain, the largest portion of the wavefront is matched for de-amplification. We note here that the waist sizes in our experiment are very close to the fundamental eigenmode-0 supported by the PSA ($>$98\% overlap for low pump powers and over 92\% at high powers). Exciting orthogonal uncorrelated eigenmodes at the input to an OPA that independently experience de-gain (or squeezing)\cite{Vasilyev2010} can circumvent mode-mixing and gain-induced diffraction\cite{ROU1995,choi1997}, which is especially critical for measuring large de-gain (or squeezing) values. By using the optimum $z$-offset, one can improve matching of the wavefront curvature to the eigenmode-0, resulting in even better overlap, especially at high pump powers. This is evident from the relatively high levels of classical de-gain observed in our traveling-wave experiment, which are within 2\,dB of the observed classical gain values (cf. Fig.\,\,\ref{fig2} inset). At such high de-gain values, phase fluctuations play an important role as indicated by the larger error bars on the high de-gain data in Figs.\,\,\ref{fig2} and \ref{fig3}. Note also that the model curves in Fig.\,\,\ref{fig2} fit the data better when a small amount of $\Delta\theta$ is taken into account via Eq.\,\,\ref{eq:nonlin.gain}. Although these measurements were done with classical fields, we anticipate that $z$-offset tuning would carry over to the quantum case, providing enhanced squeezing performance for a fundamental Gaussian mode as well \cite{muthu2011,Vasilyev2010,Koprulu1999}.

\begin{figure}[ht]
\centerline{\includegraphics[width=7.6cm]{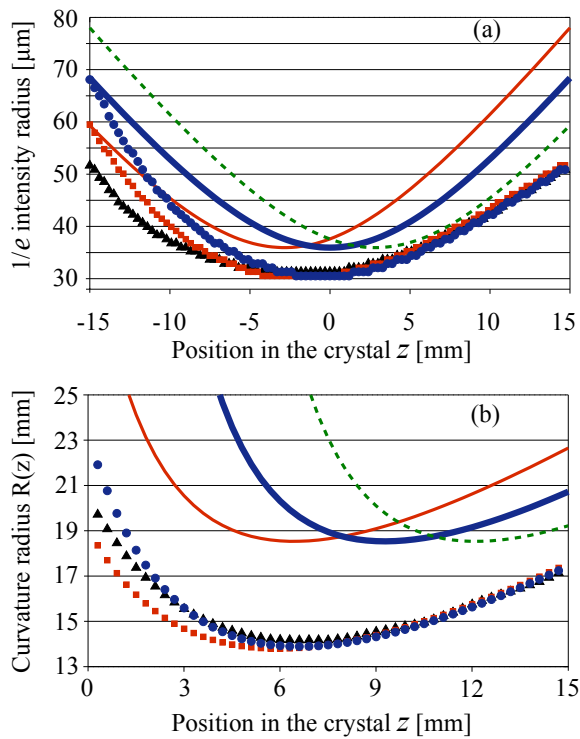}}
\caption{(Color online) Results from BPM-FFT modeling for the evolution of the signal beam's half-width (a) and curvature radius (b) in the absence (lines) and presence (symbols) of the pump with waist $a_{p}=26\,{\rm \mu}$m and power $P_{0}=450$\,W, corresponding to a PSA gain of $\sim 13.8$\,dB. Red squares and thin line are for a fundamental Gaussian input with $-2.8$\,mm offset; blue circles and thick line are for same input, but with $0$\,mm offset; green broken line is same input but with $+2.8$\,mm offset; and black triangles are the eigenmode-0.}
\label{fig4}
\end{figure}

Figure\,\,\ref{fig4} shows the modeled evolution of the signal beam's half-width $a_{s}(z)$ [Fig.\,\,\ref{fig4}(a)] and radius-of-curvature $R(z)$ [Fig.\,\,\ref{fig4}(b)] as functions of position $z$ inside the crystal. The solid lines are for the fundamental Gaussian beams of the same waists but with different $z$-offsets in the absence of parametric gain. The symbols show evolution of the same beams in the presence of a parametric pump, as well as the fundamental eigenmode-0. It is clear from Fig.\,\,\ref{fig4}(a) that parametric pumping extends the region wherein the beam width is close to the minimum (bottoms of the curves are flattened). This is because the spatially-varying gain profile narrows the signal beam's width at the earlier stages of propagation and then limits the diffractive broadening of the signal to that of the pump at the later stages. The reciprocity principle \cite{Koprulu1999} states that the eigenmode-0 profile at the crystal output should be the conjugate of that at the input. A fundamental Gaussian mode with waist $a_{s}=36\,{\rm \mu}$m and $z$-offset of $-2.8$\,mm has the best overlap (99.2\%) with the eigenmode-0 {\em at the input} (whereas the one with zero $z$-offset only has 97.2\% overlap). This means that the same mode but with an offset of $z=+2.8$\,mm has the best overlap with the eigenmode-0 {\em at the output of the crystal}. That is, to an outside viewer the eigenmode-0 at the input appears as a Gaussian beam with a waist before the crystal center, but at the output it appears as a Gaussian beam with a waist after the crystal center. Figure\,\,\ref{fig4}(b) confirms this observation by showing that the beam with $z=+2.8$\,mm offset has a curvature radius close to that of the eigenmode-0.

In conclusion, we have experimentally demonstrated and confirmed via modeling that an OPA's performance using classical fields as input can be significantly improved by judiciously tuning the $z$-offset between the waist locations of the pump and signal beams. We have also shown that with appropriate $z$-offset it is possible to approach the fundamental eigenmode of the PSA, making $z$-offset another parameter at an experimenter's disposal for optimizing the PSA performance in various applications. Such enhancement should also translate into the quantum case, where squeezed vacuum would be detected by homodyne detection with a local oscillator having the corresponding waist offset\cite{Koprulu1999}.

This material is based upon work funded by DARPA's Quantum Sensor Program, under AFRL Contract No. FA8750-09-C-0194. Any opinions, findings and conclusions or recommendations expressed in this material are those of the authors and do not necessarily reflect the views of DARPA or the U.S. Air Force.


\vspace{-4mm}
\begin{thebibliography}{10}
\newcommand{\enquote}[1]{``#1''}

\bibitem{KIM1994}
C.~H. Kim and P.~Kumar, \emph{Quadrature-squeezed light detection using a self-generated matched local oscillator}, Physical Review Letters \textbf{73}, 1605 (1994).

\bibitem{choi1997}
S.~K. Choi, R.~D. Li, C.~H. Kim, and P.~Kumar, \emph{Traveling-wave optical parametric amplifier: investigation of its phase-sensitive and phase-insensitive gain response}, Journal of the Optical Society of America B \textbf{14},  1564 (1997).

\bibitem{muthu2011}
M.~Annamalai, N.~Stelmakh, M.~Vasilyev, and P.~Kumar, \emph{Spatial modes of phase-sensitive parametric image amplifiers with circular and elliptical Gaussian pumps}, Optics Express \textbf{19}, 26710 (2011).

\bibitem{Vasilyev2010}
M.~Vasilyev, M.~Annamalai, N.~Stelmakh, and P.~Kumar, \emph{Quantum properties of a spatially-broadband traveling-wave phase-sensitive optical parametric amplifier}, Journal of Modern Optics \textbf{57}, 1908 (2010).

\bibitem{ROU1995}
R.~D. Li, S.~K. Choi, and P.~Kumar, \emph{Gaussian-wave theory of sub-Poissonian light generation by means of travelling-wave parametric deamplification}, Quantum and Semiclassical Optics: Journal of the European Optical Society B \textbf{7}, 705 (1995).
  
\bibitem{Koprulu1999}
K.~G. K\"opr\"ul\"u, and O.~Ayt\"ur, \emph{Analysis of Gaussian-beam degenerate optical parametric amplifiers for the generation of quadrature-squeezed states}, Physical Review A \textbf{60}, 4122  (1999).

\end{thebibliography}

\clearpage
\end{document}